\documentclass[11pt]{article}
\usepackage{mor-jcc,cite}

\newif\ifcameraready
\camerareadyfalse

\ifcameraready
\else
\fi

\def \MSbarbasic {\overline{{\rm MS}}}
\def \MSbar {\ifmmode \MSbarbasic \else $\MSbarbasic$\fi }

\begin{document}
\title{SUMMARY TALK I
    \ifcameraready
    \else
      \footnote{To appear in Proceedings of Rencontre de Moriond,
          ``QCD and High-Energy Hadronic Interaction'', March 22-29,
          1997, \'Editions Fronti\`eres.
      }
    \fi
}

\author{J.C. Collins}

\address{
            Penn State University,
            104 Davey Lab, University Park PA 16802, U.S.A.
}

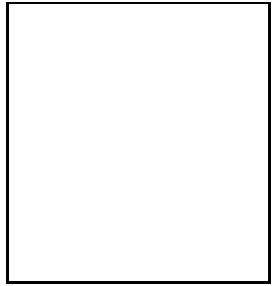
\date{6 May 97}

\maketitle

\abstracts{
    This talk covers a selection of topics that were discussed at
    this meeting in the non-heavy-ion sessions:
    (a) The high $Q^{2}$ events at HERA and their
        theoretical interpretation.
    (b) The issues of how well we know
        the predictions of QCD, with a particular emphasis on
        resummation.
    (c) Anomalies in jet production in association with $W$
        bosons at the Tevatron and in dijet production in
        deep-inelastic scattering.
    (d) Diffractive hard scattering.
}

\section{Introduction}

There were many interesting talks at this meeting, and I have
chosen to make a summary in the style of the headline news on
television: I have picked a few issues to discuss that
particularly appealed to me when I was preparing the talk.  I
apologize to the many speakers whose work I do not mention.

\section{High $Q^{2}$ events at HERA}

The most notorious results presented at this conference were
probably the excess events observed by both H1 and ZEUS at high
$Q^{2}$ and high $x$.  Since no theory talks were presented that
analyzed that data, I will give a brief summary of recent
work on the subject.  This will provide a convenient lead-in to
the rest of my talk.

\subsection{Data}

H1 \cite{H1.HQ} reported 7 deep-inelastic events in a region
where only about 1 is expected on the basis of standard model
physics.  The events appear to be ordinary DIS events, with an
electron-jet topology, and the region in which they occur is
where $M_{e-{\rm jet}}=200\pm 25{\,\rm GeV}$, $y>0.04$, and $Q^{2} >
15000{\,\rm GeV}^{2}$.

ZEUS \cite{ZEUS.HQ} reported 2 events in $Q^{2} > 35000{\,\rm GeV}^{2}$
where $0.145\pm 0.013$ are expected and 4 events in $x>0.55$ and
$y>0.25$ where $0.91\pm 0.08$ are expected.

There is a $1\%$ probability that the H1 events arise by chance, and a
few percent probability for the ZEUS events.  The backgrounds are
small, and the uncertainties in the QCD calculation of the background
are small, since normal DIS in this region is governed by valence
quark distributions.  Thus the result is a priori significant.  One
should remember the following:
\begin{itemize}

\item
    There were about 55 experimental talks presented at this
    meeting, and I estimate that each talk had between 1 and 40
    distinct physics results.  That is, many hundreds of results
    were presented, and one should expect that several results
    were presented that are significant at the $1\%$ level. 

\item
    The higher luminosity experiment, ZEUS, with $20{\,\rm pb}^{-1}$
    has a smaller signal than H1 ($14{\,\rm pb}^{-1}$).

\item 
    In assessing a small probability for any particular anomaly to
    occur, one should remember that there is a noticeable probability
    of a major bug in applications of theory and in the analysis of
    data.  It is easy to recall examples from earlier experiments.
    One should also recall Steinberger's talk here \cite{Steinberger};
    he showed how a significant anomaly in $R_{b}$ disappeared after a
    close analysis of systematic errors.

\item
    On the other hand, one should remember that the masses of 
    the muon and pion are fairly close and that the masses of the tau
    lepton and the charm quark are fairly close.
    These coincidences between the
    masses of first and second generation hadrons and certain leptons
    caused confusion in the initial investigation of these particles.
    Perhaps a leptoquark of mass 200 GeV is there to create equal
    confusion in the investigation of the slightly less massive top
    quark?  

\end{itemize}

Clearly, one cannot say that there really is new physics until the
results have withstood the scrutiny of (a) a repetition of the
measurements in a new run at HERA, (b) a search for corresponding
physics (see Sect.\ \ref{High.Q.Theory}) at other places (Tevatron and
LEP), and (c) a careful check for errors in both theory and experiment.
Too great a scepticism is also unwarranted. One must treat searches
for new physics in the same way as a drug company's search for
compounds with interesting pharmaceutical properties.  A first round
of screening of a large collection of compounds will inevitably
produce false positive results.  A second round of tests is necessary
to verify which are the real results.

The cautionary tale of the discovery of the $J/\psi $ must also be
remembered.  The $J/\psi $ was first seen \cite{First.DY} as what we
now know is a substantial shoulder above the Drell-Yan continuum.
The shoulder was the narrow resonance smeared out by detector
resolution.  But one of the conclusions of the paper was ``No
resonances (i.e., $1^{-}$ bumps) are observed, \dots''.  In
fairness to the experimentalists we should remember that the
theorists weren't doing so well: the Drell-Yan curve was well
above not only the continuum but also the smeared resonance!
Despite this, we now know that the Drell-Yan model provides a
correct first approximation to real QCD predictions.  We clearly
see how much theory has improved in the last 25 years.

\subsection{Theory}
\label{High.Q.Theory}

A consistent picture of how to interpret the HERA excess appears
in a large number of recent preprints \cite{HERA.Phen}.  None of
this work was reported at this meeting, although a number of
relevant experimental results were reported.

New physics below its threshold can be described by a contact
interaction.  The obvious contact interaction needed to get an excess
of DIS events is a 4-fermion electron-quark vertex.  To get the
correct number of events, this must be stronger than the standard
electro-weak interaction due to $Z$ and $\gamma $ exchange, in the
relevant kinematic range.  Such an interaction must give substantial
effects at LEP.  There are several possible Dirac-matrix structures,
and current data rule out almost all relevant terms in a contact
interaction.  With more statistics, all the terms can be ruled out.

It is therefore more reasonable to postulate an interaction with an
enhancement in the electron-quark channel.  This can be most easily
modeled as the production of a leptoquark, either of spin zero or spin
one. Its mass should be about 200 GeV, according to the H1 data.  The
value of the electron-quark-leptoquark coupling $\lambda $ depends on
which quark is involved and on the branching ratio to the observed
channel. If one assumes a branching ratio of one and a coupling to the
predominant valence quark $u$, then one gets the smallest coupling,
substantially smaller than the electromagnetic coupling (0.02 to
0.05).  This implies that the leptoquark is a narrow resonance.

Since a leptoquark must have normal QCD couplings to gluons, it
can be pair-produced at the Tevatron.  In the case of a scalar
leptoquark, the cross section is completely predicted.  In the
case of a vector leptoquark there is an anomalous color magnetic
moment coupling, and there is a range of predictions for the
cross section, with the lower bound being higher than the cross
section for a scalar leptoquark.  At this meeting, Valls
\cite{D0.leptoquark} showed the bounds from the Tevatron: 175 GeV
for a scalar leptoquark with a 100\% branching ratio to an
electron plus quark, and 298 GeV for a vector leptoquark.
Because of the weak value of the coupling, the rate for single
leptoquark production ($g+q \to  e + {\rm leptoquark}$) is too small
to be measured.

So a leptoquark is not quite ruled out.

Clearly we should have better data within a year.  The HERA
experiments will be able to confirm the excess, and if the excess
is real, then there are predictions for LEP and the Tevatron that
it will be possible to test.

\section{QCD}

Overall, we see good agreement between experiment and the
predictions of QCD.  (In this context, this means perturbative
predictions, since these are the only predictions we have for
high-energy scattering.)  A good example (out of many
possibilities) is a plot shown by Ghez \cite{Ghez} of event shapes
measured by the OPAL collaboration in $e^{+}e^{-}$ annihilation at 161
GeV.  Given the value of $\alpha _{s}$ (at some given scale in a given
scheme), the predictions are absolute.  Basically, the
predictions are obtained using fixed-order NLO calculations.  But
at the more extreme values of the event-shape variables
resummation is needed to get accurate predictions.

The result of much experience is that we trust QCD as a
predictive theory of strong interactions.  (However, we do not
trust ourselves to get these predictions correct every time!)

A larger class of cross sections involve non-perturbative
quantities: parton densities, fragmentation functions and their
kin.  The predictive power of QCD is, of course, that the parton
densities, etc., can be measured in a limited set of processes
and then used to predict many other processes. One example is the
single-jet-inclusive cross section at the Tevatron.  Nang's talk
\cite{Nang} showed data from CDF and D0.  The D0 data is in
agreement with the predictions of a range of jet $E_{T}$ from 50 to
500 GeV, while the cross section falls by close to 7 orders of
magnitude.  But it should be noted that the errors get quite
large at the higher values of $E_{T}$: up to 50\%.  As is
well-known, the CDF data becomes somewhat high at large $E_{T}$.
Despite this, the CDF and D0 data are in agreement.  (This is no
contradiction since equality within errors is not a transitive
relation, i.e., if $A=B$ and $B=C$ within errors, it does not
follow that $A=C$ within errors.)

If one wants to know where the excess events in D0 disappeared
to, see Sect.\ \ref{W.jets}.

\subsection{Errors}

This brings me to some important issues:
\begin{itemize}

\item How well do we know parton densities?  I.e., what are the
    errors on the global fits?

\item How accurately can we make QCD predictions?  I.e., how
    accurate are the theoretical formulae?

\end{itemize}
It should be obvious, for example from many talks at this meeting,
that QCD predictions are critical to new physics searches.  Signals of
new physics are generally a difference between data and a
standard-model prediction with a large QCD components.  Clearly we
need to have a reliable quantitative estimates of the errors in our
QCD predictions.  This is quite non-trivial.

For example, predictions for cross sections at the Tevatron and at
HERA rely heavily on the values of parton densities.  These are
obtained from global fits that involve 30 to 40
parameters in the theory and over a thousand data points from
many different experiments.  The difficulty of obtaining reliable
errors can be illustrated by observing that adjusting a correctly
defined $\chi ^{2}$ to within 1 unit of its minimum is the correct way
of obtaining the best fit, whereas the actual value of $\chi ^{2}$ will
be about 1000.  Thus $\chi ^{2}$ must be calculated to better than
0.1\% accuracy.

In the case of purely statistical errors, we understand $\chi ^{2}$
and have no problem understanding its calculation.  But in many cases
systematic errors dominate.  Almost by definition, we do not
understand the statistics of systematic errors, so any attempt to
treat them suffers from some imprecision.  But we must try to do our
best.

Perhaps the most important feature of systematic errors in a
measurement of a differential cross section is that they are
normally highly correlated between different data points.
Consider a measurement of a structure function $F_{2}(x)$ with, say,
12 data points when the dominant systematic error is a
calibration error with a definite shape in $x$.
A 1 $\sigma$ change in the data due to the calibration error is not
very significant, and if $\chi^2$ is correctly calculated it should
only change by about 1 unit due to the change in the data.  But if we
were to follow the common procedure of adding systematic and
statistical errors in quadrature, then a deviation between theory and
experiment of such a form would change the calculated $\chi ^{2}$ by
12 units --- enormously significant, apparently.

Unfortunately in global analyses, incorrect calculations of the
effect of systematic errors on $\chi ^{2}$ are typically used.  This is
not through incompetence, since the necessary information is not always
available.  This situation has to be changed.  Without a correct
treatment of correlated systematic errors, the effects of certain
deviations between theory and experiment are enormously
overstated.  If one consequently adjusts the threshold in $\chi ^{2}$
which one considers to give a significant deviation between
theory and experiment, there is a danger of ignoring many
genuinely significant deviations.

Although error analysis appears to be a recondite subject, it is vital
to the progress of QCD and to the search for new physics.  So it was
very gratifying to hear Yu's talk \cite{Yu} about an analysis of
neutrino scattering data from CCFR and other experiments with a
correct treatment of correlated errors.  Another improvement in the
accuracy of this work was the use by Kataev and collaborators
\cite{Kataev} of NNLO calculations.  A combination of the methods in
these two talks will result in very reliable determinations of
$\Lambda _{\MSbar}$ and of the non-singlet quark densities.

I should add that good analyses should attempt to treat errors on
the theory due to higher order corrections, etc., in a similar
fashion.  The errors are highly correlated.

\subsection{Resummation}

Unfortunately, the theoretical errors are often bigger than the
experimental errors.  This is very embarrassing to theorists, of
course, and work is urgently needed to remedy the situation.
Brute force higher order calculations are only part of the
answer, because higher order corrections with the standard
techniques perturbative QCD are often excessively large.  One can see
the symptoms of this in many of the talks at this meeting.

One reason for the large corrections is that the cancellation of
soft and collinear parton emission is not particularly good.  For
example, one might meet a LO and an NLO term like
\begin{equation}
   f(x) + \frac {\alpha _{s}}{2\pi } \int _{x}^{1} d\xi  \, f(\xi )
          \left( \frac {1}{\xi -x} \right)_{+} .
\end{equation}
There is a divergence in the integral over the region $\xi >x$, and
the plus-distribution cancels this by what is in effect a
delta-function with an infinite coefficient localized at $\xi =x$.
If the parton density $f(x)$ is rapidly varying, this
cancellation is numerically poor and results in a large
numerical coefficient for the $\alpha _{s}$ term.

There are many variations on this theme.  The problem is not just that
one gets a large NLO correction, but that a typical source of the
large correction will also induce large corrections in even higher
order terms.  The only remedy for this situation is to reorganize the
theory, for example by a resummation over all orders of the large
terms.  Only by such means does one get accurate predictions.  

Indeed, if an appropriate resummation is not done, one can easily end
up with unphysical cross sections: for example an isolated direct
photon cross section that is larger than the inclusive cross section.
There were a number of talks in this area, and it is evident that
resummation is quite essential in many processes.  Experimental
evidence was quite nicely shown in a plot of event shapes at SLD.
Without resummation, theory (at $O(\alpha _{s}^{2})$) fails to
describe, for example, the thrust distribution below a thrust of about
0.1.

However, doing the resummation correctly is somewhat tricky. For
example, while Berger \cite{direct.photon.B} and Fontannaz
\cite{direct.photon.F} agreed on the need for resummation in the cross
section for the production of isolated direct photons in hadron-hadron
collisions, they disagreed on the details, including such a
fundamental question as to whether the cross section is finite before
resummation. The kinematics of the soft gluons that cause the problem
is tricky enough that it takes a substantial time for outsiders (like
myself) to understand the issues well enough to gain a reliable
opinion on the rights and the wrongs of the issue.

But getting these details sorted out will be essential to the
progress of QCD and of its applications.  Resummation in all its
varieties is essential to high-precision phenomenology.

\section{Anomalies in jet cross sections}

There were two anomalous results presented at this meeting that
seemed to represent more significant problems than the widely
publicized large $Q^{2}$ events at HERA and the high $E_{T}$ jets at
CDF.

\subsection{$W+ {\rm jets}$ at D0}
\label{W.jets}

The first \cite{Dittmann} was in the rate of $W+{\rm jet}$ events in
the D0 experiment.  Between 20 and 60 GeV, the ratio $\sigma (W+1\ 
{\rm jet}/\sigma (W+0\ {\rm jets}$ is consistently 50\% and more above
theory, well outside the errors on both theory and experiment.  This
is in a domain where one generally considers QCD and the standard
model to be valid.  In particular, inclusive $W$ cross sections and
inclusive jet cross sections are in agreement with QCD predictions.

For the same cross section, data was presented \cite{Dittmann}
from the CDF experiment, and the graphs show good agreement
between theory and experiment.  However, this good agreement only
concerns the shape of the cross section, as a function of jet
$E_{T}$.  The theory curve was normalized to the data.

\subsection{Dijets in DIS}

A second deviation between theory and experiment was in dijet
rates in DIS \cite{Spiekermann}.  H1 has found that the rate of
dijets in DIS is consistently well above NLO QCD predictions for
a whole kinematic range ($5 < Q^{2} < 100 {\,\rm GeV}^{2}$ and $10^{-4} <
x < 10^{-2}$). Earlier measurements \cite{H1.multijets} of multi-jet
rates were in agreement with QCD models. This gives a ``problem
for [the] measurement of $\alpha _{s}$ or [the] extraction of [the] gluon
density from jets''. Interestingly, the color dipole model
(ARIADNE) does describe the data perfectly well.  Since the
values of $Q^{2}$ and $x$ are rather small, it is likely that
the problem is that the use of conventional fixed order
calculations that is wrong.  (Note in particular that $Q^{2}/s$ in
this data sample goes down to $5\times 10^{-5}$.)  
It is the higher
$Q^2$ data that should be used for measuring $\alpha_s$ with the aid
of fixed order QCD calculations.

This point is further strengthened when one observes that the new data
involves a minimum $p_{t}$ for the jets of 5 GeV, which is rather
larger than the scale $Q$ set by the virtual photon. A treatment in
analogy with photoproduction is perhaps appropriate, and on the basis
of the $x_{\gamma }$ distribution, the speaker suggested that there is
evidence for a resolved component in virtual photons.  (I would add
that since the photons have a $Q^{2}$ that is in the perturbative
region, this resolved component should be amenable to a perturbative
analysis.)

Clearly more theoretical work is needed here, and it is not
obvious that we have a contradiction with QCD.  Certainly it is an
area ripe for the application of suitable resummation techniques.

\section{Diffraction}

Over the past few years we have seen a renaissance of interest in
diffractive scattering.  It is now respectable to discuss the
pomeron.  It is not as if the physics represented by the pomeron
ever disappeared; it gives, in fact, the bulk of hadronic cross
sections.  An important but extremely difficult problem is to
understand pomeron physics, diffraction in particular, within
QCD.  What is possible for the first time in the current
generation of experiments is to study hard scattering in
association with diffraction.

Diffractive processes can be defined as those with non-exponentially
suppressed rapidity gaps.  There are several distinct classes of
diffractive processes that are being investigated: (a) those with
rapidity gaps between a beam hadron and the rest of the final state,
(b) those with rapidity gaps between jets, and (c) highly exclusive
final states in photo- and electro-production.  Because of limited
time I only discussed the first of these topics.

We saw the latest data from CDF, D0, H1 and ZEUS
\cite{CDF.diff,D0.diff,HERA.diff}.  This may be regarded as
probing the partonic structure of the pomeron, with the
Ingelman-Schlein model representing a benchmark model.

One interesting property is that diffractive hard scattering is
much rarer in Chicago than in Hamburg: the fraction of hard
scattering that is diffractive is typically under 1\% in
hadron-hadron collisions at the Tevatron, whereas the fraction is
closer to 10\% at HERA.  Thus it has taken rather longer for CDF
and D0 to establish their signals for diffractive hard
scattering.

As Kaidalov \cite{Kaidalov} emphasized, an exchanged pomeron is not to
be regarded as a real particle.  Its parton densities do not obey a
momentum sum rule. Moreover, there are absorptive corrections in
diffractive hadron-hadron scattering, but not in diffractive DIS.
This implies that a simple application of the methods perturbative QCD
will not predict the Tevatron cross sections given the ones at HERA,
and that the cross sections should be rather lower than predicted.
This presumably explains the lower diffractive fractions at the
Tevatron.

Quantitative phenomenology is progressing very nicely at HERA.  Given
the absence of the absorptive corrections, one can treat the results
as directly measuring the parton densities of the pomeron.  (Though
whether this is legitimate is another question.)  H1 showed for the
first time data on diffractive dijet production in DIS.  The measured
cross sections can only be fit if there is a large gluon content in
the pomeron, since the quarks are constrained by diffractive DIS.
This very prettily confirms the large gluon content deduced by ZEUS
from their diffractive photoproduction data (and also confirms that
the theoretical picture is consistent).  It is also consistent with
the results obtained by H1 by a quantitative analysis of the scaling
violations in their data on $F_2^{\rm diff}$. We must regard the
pomeron as a glue-ball state to a first approximation.

With these analyses it is clear that the Tevatron data are well below
the expectations if hard-scattering factorization holds.  Goulianos
\cite{CDF.diff} showed a plot of momentum fractions in the pomeron. At
both the Tevatron and at HERA we have measurements of two or more
processes, so that both the quark and the gluon content of the pomeron
are constrained.  The Tevatron cross sections are a factor of several
below those that one would expect on the basis of the HERA.  This is
very direct evidence for absorptive corrections, and therefore
diffractive hard scattering is proving to be an effective tool as a
microscopic probe of the structure of soft hadronic cross sections.

We can expect more incisive results in the future, for example,
from an analysis of the data \cite{HERA.diff} from ZEUS where the
diffracted proton is explicitly measured (and corresponding future
data from the other experiments).

\section{Conclusions}

\paragraph{Clich\'e 1}
There are a few anomalies among the generally good agreement
between theory and experiment.  The notable ones are: (a) the high
$Q^{2}$/high $x$ events at HERA, (b) an excess of $W + {\rm jets}$
events in D0, (c) an excess of dijet events in DIS in H1.  The
lasts two are in an area in which we expect QCD to be valid.  The
first is in an extreme domain of $Q$, so is a potential signal of
new physics; but the new physics should be visible in other
experiments at LEP and/or the Tevatron in the near future.

\paragraph{Clich\'e 2}
We have seen a number of significant effects disappear or not
find themselves confirmed by other experiments: the $R_{b}$ anomaly
at LEP, the four-jet events at ALEPH, the high $E_{T}$ jets at CDF.
Given the success of the standard model, one must put the balance
of the probability on other anomalies as being due to a
fluctuation or an error of some kind.

\paragraph{Clich\'e 3}
QCD is a highly non-trivial theory, as are the experiments needed
for its investigation.  A high level of quantitative
understanding of QCD is needed in current and future searches for
new physics.  There are many prospects for gaining understanding
of QCD in innovative ways: e.g., the various programs on
resummation, and the active work on diffractive physics.  Data
that is improved in accuracy and in the range of phenomena
measured is pushing the theorists to match the experiments.

\section*{Acknowledgments}

My work is supported in part by the U.S. Department of Energy
under grant number DE-FG02-90ER-40577.  I would like to thank CERN and
DESY for their hospitality while this paper was written.


\end{document}